\documentclass[pra,aps,twocolumn,superscriptaddress,nofootinbib,letterpaper,balancelastpage,notitlepage,floatfix]{revtex4-2}
\usepackage{graphicx}
\usepackage{bm}
\usepackage{braket}
\usepackage{siunitx}
\usepackage{hyperref}

\begin{document}


\title{Heading Error Compensation in a Portable Optical Magnetometer Using a Double-Pass Single Beam Configuration}

\author{Y. Rosenzweig}
\affiliation{Elta Systems Ltd, Land Division}
\affiliation{Department of Physics, Ben-Gurion University of the Negev, Israel}

\author{D. Tokar}
\affiliation{Elta Systems Ltd, Land Division}

\author{I. Shcerback}
\affiliation{Elta Systems Ltd, Land Division}

\author{M. Givon}
\affiliation{Department of Physics, Ben-Gurion University of the Negev, Israel}

\author{R. Folman}
\affiliation{Department of Physics, Ben-Gurion University of the Negev, Israel}


\begin{abstract}
Optically pumped magnetometers are ultra-sensitive devices, but this sensitivity can significantly degrade due to heading errors, whereby a change in the angle between the pumping laser and the magnetic field translates to a change in the magnetic field readout. We present a portable all-optical single-beam magnetometer with a reduced heading error due to a double-pass configuration. We analyze it both theoretically and experimentally.
In addition to this significant improvement in performance, the increased interaction length of the laser with the cell enhances the signal. Overall, the new configuration enables better accuracy, as well as the reduction of the cell temperature, laser power, and further miniaturization of the sensing head. This work opens the door for a simple and robust sub-pT portable sensor in Earth field.     
\end{abstract}
\maketitle

\section{Introduction}\label{Sec: into} 
An optically pumped magnetometer (OPM) can measure magnetic fields with ultra-high sensitivity. The recent progress in the field during the 21st century with the invention of the Spin-Exchange-Relaxation-Free (SERF) zero-field magnetometer\,\cite{allred2002high} outperformed even the superconducting quantum interference device (SQUID) by setting the experimental sensitivity limit of the SERF to 160\si{\atto\tesla}\,\cite{dang2010ultrahigh}. The same research group also demonstrated a sub-fT sensitivity in a low non-zero magnetic field using a multi-pass configuration\,\cite{sheng2013subfemtotesla}. Combining the OPM sensitivity with its relatively low cost made the SERF OPM a strong candidate to replace the expensive and bulky superconducting-based magnetometers when sensing extremely low magnetic fields such ones evolving from brain activity\,\cite{pratt2021kernel}, heart\,\cite{shah2013compact} or even exotic fields\,\cite{wang2018application}. Nonetheless, when the bias field is as high as Earth's magnetic field, the actual sensitivity of a portable Earth field magnetometer is around 1\,pT at 1\,Hz\,\cite{lucivero2014shot,acosta2006nonlinear,oelsner2022integrated,limes2020portable}, mainly due to the non-linear Zeeman effect (NLZ) broadening the magnetic resonance line\,\cite{li2016unshielded}.
\par
While the OPM is considered a scalar magnetometer, under Earth's magnetic field, it is not invariant under rotation due to the heading error: When the magnetometer is placed in a large magnetic field, such as Earth's magnetic field, the Zeeman effect can no longer be considered linear and different Zeeman levels have different Larmor frequencies. In addition, the ground state population distribution, under circularly polarized pumping light, strongly depends on $\theta$- the angle between the magnetic field and the laser propagation vector. As will be explained in details later on, the combined effect of the NLZ effect with different ground state population distribution at different angles results in a change of the magnetic field readout. The combination of NLZ effect and different population distribution at different $\theta$ angles is the main contribution to the heading error effect, although there are other minor contributions\,\cite{zhang2023heading}. Another important contribution is the light shift, especially in the case of off-resonance pumping\,\cite{schultze2017optically}, but it is irrelevant for an all-optical magnetometer where the AC Stark effect from the off-resonance pumping induces pseudo-magnetic field which oscillates near the Larmor frequency (and far above the magnetometer bandwidth), replacing the need for a microwave field\,\cite{bell1961optically}. 
Under Earth's field, the heading error can be as high as a couple of dozen nT\,\cite{hrvoic2005brief}, effectively masking the actual OPM sensitivity when attached to a portable platform. 
\par
In this work, we present a method to mitigate the effect of the heading error. While several other methods have been developed to this end, such as split-beam configuration\,\cite{yabuzaki1974frequency}; adding a secondary modulation at the revival frequency\,\cite{seltzer2007synchronous}; adding an RF field to spin-lock the atoms\,\cite{bao2018suppression}; alignment-based magnetometery\,\cite{zhang2023heading} etc., most of them, while performing well in the lab has not been implemented in a commercial portable platform. To our knowledge, only the split-beam can be found in a commercial OPM\,\cite{gem, scintrexltd}, and although conceived first in 1974\,\cite{yabuzaki1974frequency} active research using the concept of the split-beam is ongoing\,\cite{oelsner2022integrated, schultze2017optically}. We show that a double-pass beam configuration in a portable magnetometer can significantly attenuate the heading error while exhibiting a higher signal-to-noise ratio (SNR) than the split-beam configuration. In addition, doubling the interaction length boosts the sensor's signal, power budget, and miniaturization. All critical factors in a portable Earth field magnetometer\,\cite{hovde2013commercial}. 
\par
In a double-pass configuration, a circularly polarized laser light traverses the cell and is reflected to the cell without spatial overlapping while keeping its helicity. The heading error from the incoming beam is equal to that of the reflected beam but with an opposite sign due to the reflection symmetry of the heading error (as will be shown later on). As the laser passes the cell, it sums the signal from the transmitted and reflected paths, resulting in an unshifted signal. This results in a reduction of the heading error. In addition, no expensive optical elements, such as additional laser, beam-splitter, specially designed mirrors, or $\lambda/4$ retarder, which are needed in other heading error reduction methods, are required here. 
\par
The rest of the paper is organized as follows: in Sec.\,\ref{Sec: Theory}, we will explain the angle dependency on the magnetic field readout and how a double-pass configuration can address this problem. Then, in Sec.\,\ref{Sec: experimental}, we will present our portable magnetometer and its performance regarding the heading error; in Sec.\,\ref{Sec: conclusion}, we will summarize the paper.

\section{Theory}\label{Sec: Theory}
\begin{figure}[ht]
\centering
\includegraphics[width=1\columnwidth]{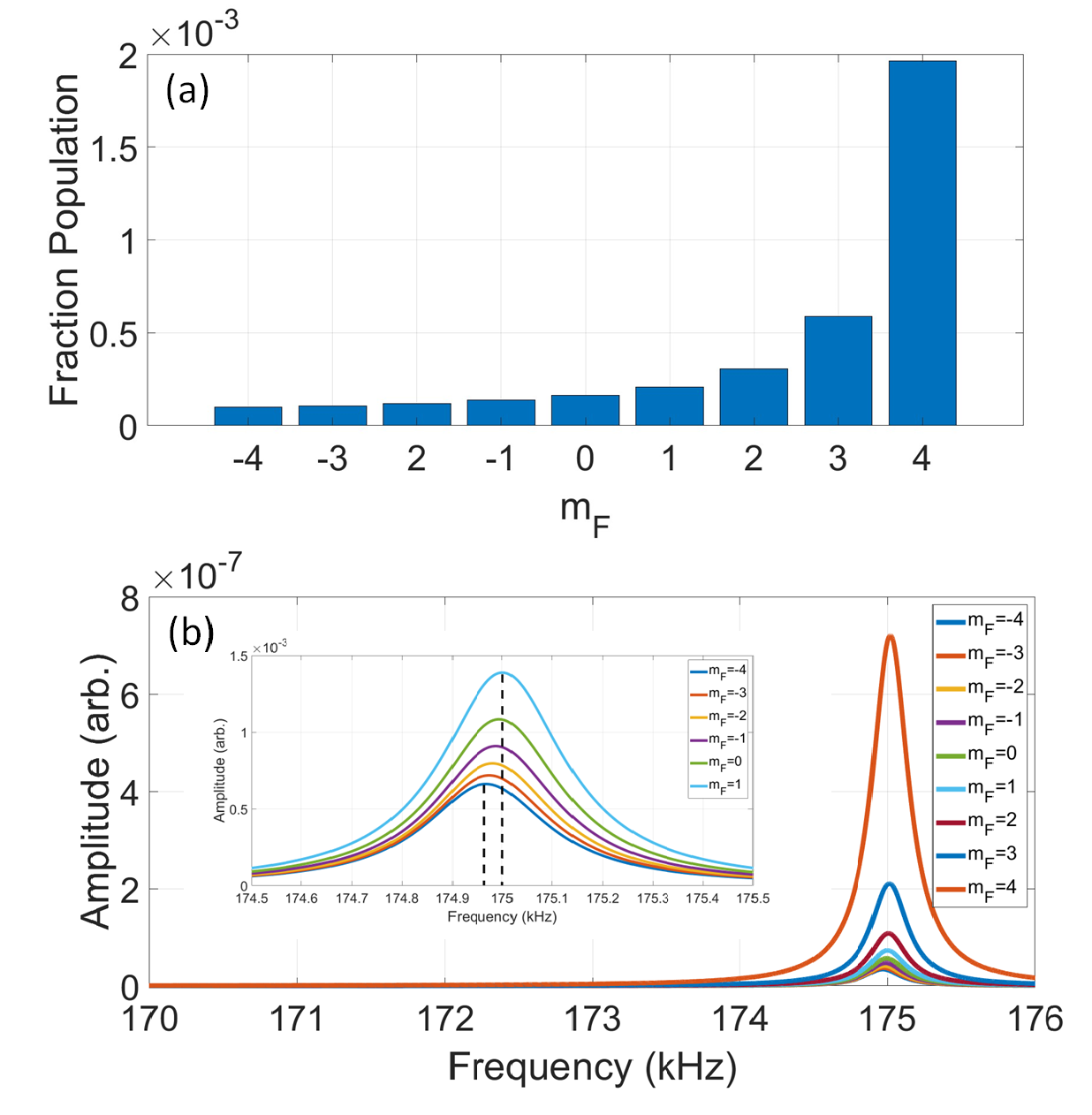}
\caption{(a) The population distribution at the different Zeeman levels of $F=4$ due to optical pumping from $F=4$ to $F'=3$ and relaxation to $F=3,4$. The laser power was set to $100\,$\si{\mu\watt} and $\theta$, the angle between the laser propagation vector and the magnetic field, was set to $60\,$\si{\degree}. (b) The resonance line shape of the different Zeeman levels according to Eq.\,\ref{eq: magnetic resonance}. In the inset, we zoom in to better demonstrate the shift in the mean Zeeman frequency of each $m_F$ level due to the NLZ effect, and a dashed black line from the center of $m_F=-4$ and $m_F=1$ was added to clarify the shift in the resonance frequency. Different population distributions in (a) due to a change in $\theta$ will result in different amplitudes for the resonance signals in (b), causing the peak of the summed signal to change accordingly, which is the primary mechanism behind the heading error.}
\label{fig: population and resonance}
\end{figure}

\begin{figure}[ht]
\centering
\includegraphics[width=1\columnwidth]{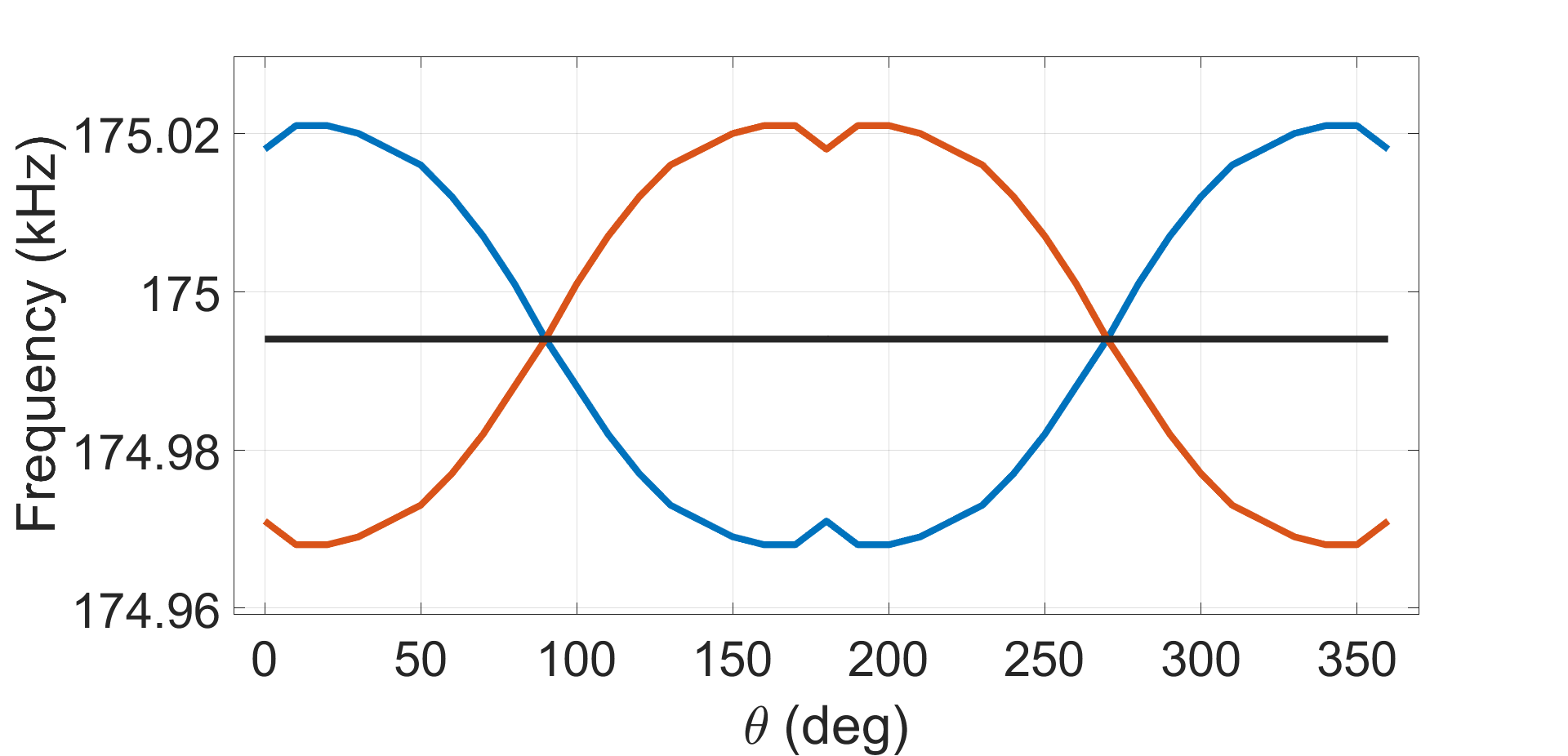}
\caption{Magnetic resonance frequency readout as a function of $\theta$. Left (right) circular polarization in red (blue). The resonance is extracted from the maximum of the signal in Eq.\,\ref{eq: weighted signal}, and the Larmor frequency and population of each $m_F$ state were calculated with the same parameters as in Fig.\,\ref{fig: population and resonance}. The black line shows the magnetic resonance frequency extracted from the subtraction or summation of the signals calculated using Eq.\,\ref{eq: weighted signal} for left and right circular polarization (for subtraction, we extract the resonance frequency from the zero crossing of the signal, instead of the maximum, since the resulted signal of the subtraction is a dispersive line shape). We can see that a summation or subtraction of a magnetic resonance signal due to the pumping beam's left and right circular polarization will cancel the angle dependency on the resonance frequency for any angle. Notice that the maximal/minimal heading error is not at $\theta=0$\si{\degree} or 180\,\si{\degree}. As explained in the text, this is because when a circularly polarized laser is tuned to the transition between $\ket{F=4}$ and $\ket{F'=3}$ with $\theta=0$\,\si{\degree}, states $m_F=3,4$ are dark states. But, as $\theta$ starts to change above/below zero, then $m_F=3$ is no longer a dark state, further pumping the population towards $m_F=4$, and as a consequence, the resonance frequency readout will be higher. Further increasing $\theta$ will result in stronger $\sigma^-$ transitions pushing the population towards lower resonance frequencies. It is clear from the symmetry of the results that the change in the magnetic resonance frequency readout can be eliminated by averaging or subtracting the two signals.}
\label{fig: heading error simulation}
\end{figure}

In order to illustrate the impact of $\theta$ on the magnetic field readout, we start by calculating the steady-state Zeeman distribution, $n_{m_F}$, under optical pumping from a circularly polarized light using rate equations. We assume a room temperature Doppler broadened $^{133}$cs with ground state $\ket{F=4, m_F}$ and $\ket{F=3, m_F}$ and excited state $\ket{F'=3, m'_F}$. Following \cite{oelsner2019sources}, we calculate the population distribution among the different levels using rate equations in which the levels are coupled due to the absorption rate, $w$, and the relaxation rate $B$. The absorption rate can be expressed using the Fermi golden rule

\begin{equation}\label{eq: absorbption rate}
    W_{m_F,m_{F'}}=\frac{2\pi}{\hbar}\left\vert\bra{4,m_F}\vec{D}\cdot\vec{E}\ket{3,m_{F'}}\right\vert^2\int_0^\infty \rho(\omega)s(\omega)d\omega\,,
\end{equation}
where $\vec{D}$ is the dipole operator, $\vec{E}$ is the electric field of a circular polarized laser, $\rho$ is the laser line shape and $s$ is the optical transition line shape. The relaxation rate, $B$, can be expressed as\,\cite{steck2007quantum}
\begin{equation}\label{eq: relaxation rate}
    B_{m_F,m_{F'}}=\frac{2\omega^3_{opt}}{3\epsilon_0hc^3}\left\vert\bra{4,m_F}\vec{D}\ket{F,m_{F'}}\right\vert^2\,,
\end{equation}
where $c$ is the speed of light, $\epsilon_0$ is the permittivity and $\omega_{opt}$ is the optical resonance frequency. In addition, we added a thermal relaxation $T_1=10$\,ms between all ground state levels, and added an additional equation in order to normalize the population (explicitly: $\sum_{m_F=-F}^{m_F=F}n_{m_F}=1$ where $n_{m_F}$ is the population at level $m_F$, summed over the hyperfine levels $F=3,4$ and $F'=3$). We solve the 24 steady-state rate equations and find the population in each state. Once we know the fraction of the population in each state, we calculate the magnetic resonance frequency for the $F=4$ Zeeman states\,\cite{bao2018suppression}
\begin{equation}\label{eq: non linear resonance}
    \omega_{m_F}\approx \frac{\mu_BB}{4\hbar}+\frac{(\mu_B B)^2}{16\hbar\Delta_{hf}}(2m_F-1)\,,
\end{equation}
where $\mu_B$ is the Bohr magneton, $B$ is the magnetic field and $\Delta_{hf}$ is the hyperfine splitting. Each Zeeman level has a distinct population, $n_{m_F}$ and a resonance frequency $\omega_{m_F}$. Assuming a typical magnetic resonance line shape in the form of a Lorentzian\,\cite{kimball_2013}, we find for each level its associate line shape
\begin{equation}\label{eq: magnetic resonance}
L_{m_F}=n_{m_F}\frac{1}{\pi}\frac{0.5\Gamma}{(\omega-\omega_{m_F})^2 + (0.5\Gamma)^2}\,,
\end{equation}
where $\Gamma$ is the decoherence rate, which was taken to be $500$\,\si{\hertz} in our calculation, and $\omega_{m_F}$ is the magnetic resonance calculated by Eq.\,\ref{eq: non linear resonance} for $B=50\,$\si{\mu\tesla}. The population of all the nine Zeeman sub-levels of $F=4$, and their associated magnetic line shape is presented in Fig.\,\ref{fig: population and resonance} for pumping power of $100\,\mu W$ and $\theta=60$\,\si{\degree}. The observed signal is the sum of all the Zeeman sub-levels' magnetic line shape
\begin{equation}
\label{eq: weighted signal}
L_{tot}=\sum_{m_F=-F}^{F} L_{m_F}\,.
\end{equation}
The magnetic resonance frequency is extracted from the maximum of $L_{tot}$. In Fig.\,\ref{fig: heading error simulation}, we calculate magnetic resonance readout (i.e., the maximum of $L_{tot}$) for different angles for both left and right circular polarization, and we can see how the magnetic field readout is changed due to a change in $\theta$. Also, notice the symmetry for opposite circular polarization, with a maximum difference of $53$\,\si{\hertz} (15\,\si{\nano\tesla}). This value is consistent with the estimation mentioned in ref.\,\cite{hrvoic2005brief}: $\approx$20\,nT for Cs.
\par
The maximal and minimal heading error in Fig\,\ref{fig: heading error simulation} is slightly above and below $0$\,\si{\degree} and $180$\,\si{\degree} for left/right circular polarization. This is due to the fact that for circularly polarized light, on resonance with $\ket{F=4}$ to $\ket{F'=3}$, both $m_F=3,4$ are dark states for $\theta=0$\,\si{\degree} as only $\sigma^+$ transition is allowed. When $\theta$ starts to deviate from $0$\,\si{\degree}, a small component of $\sigma^-$ and $\pi$ transitions are introduced, and state $m_F=3$ is no longer a dark state. Thus, effectively further pumping the population into state $m_F=4$, which results in an up-shift of the average resonance frequency due to the $m_F$ dependency in B, as was shown in Eq.\,\ref{eq: non linear resonance}. If $\theta$ is further increased, the contribution of $\sigma^-$ transition is more dominant, pushing the population into negative values of $m_F$, and the resonance frequency begins to drop in a sine-wave like behavior, and vice-versa for $\theta=180$\si{\degree}.
\par
\begin{figure}[h]
\centering
\includegraphics[width=1\columnwidth]{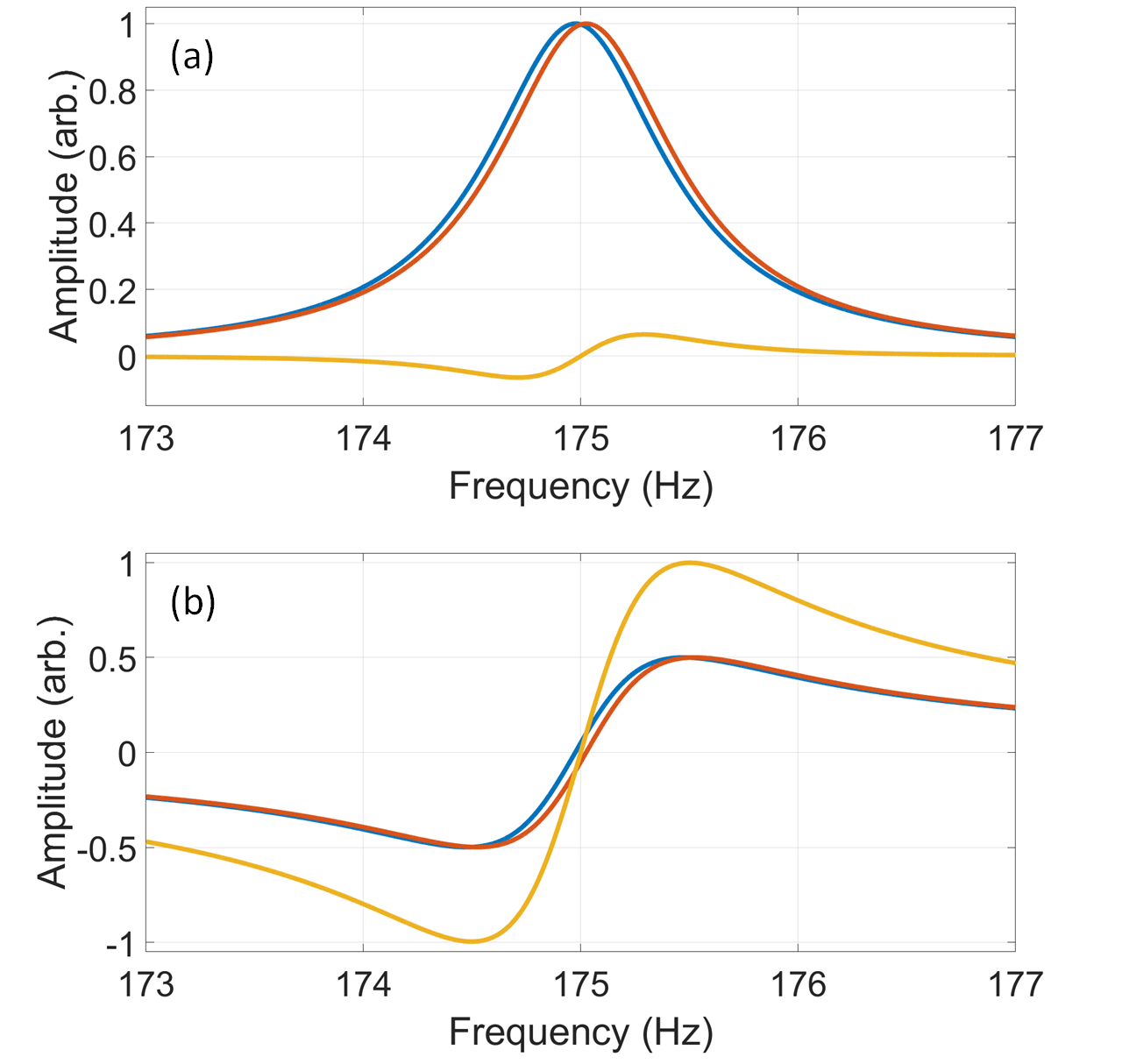}
\caption{(a) Split-beam signal: Two in-phase magnetic resonance signals (red and blue line) separated by $50\,$Hz and their subtraction (yellow), as a function of the pumping rate modulation frequency. (b) Double-pass signal: Two quadrature magnetic resonance signals (red and blue line) separated by $50\,$Hz and their summation (yellow), as a function of the pumping rate modulation frequency. The in-phase and quadrature line shapes were taken from Eq.\,\ref{Eq: bloch Mx} with $\Gamma=500\,$Hz, and their amplitude was normalized. The $50\,$Hz difference in resonance frequency (i.e., heading error) and the $500\,$Hz relaxation rate are typical Earth field Cs magnetometer values. For the above values, we see that the slope is much stronger for the double-pass configuration, making it a much more sensitive method.}
\label{fig: u and c shape}
\end{figure}

In order to emphasize our suggested method's advantage, we start with a short description of the common split-beam configuration: A linear-polarized beam is split into two parallel beams before entering the vapor cell. After the splitting, one beam is polarized using a retarder to right-hand circular polarization and the other to left-hand circular polarization. After the beams traverse the cell, the signal is subtracted using a balanced photo-diode, and the magnetic field is extracted. Due to the symmetry between left/right circular polarization shown in Fig.\,\ref{fig: heading error simulation}, the subtracted signal will have a reduced heading error\,\cite{yabuzaki1974frequency}. In a double-pass configuration, on the other hand, there is no need to split the beam, polarize them separately, or use a balanced photo-diode. Instead, a circularly polarized single beam traverses the cell and reflects back to the cell (with no spatial overlapping with the transmitted beam), keeping its helicity. The beam sums the contribution to the signals from $\theta$ (transmitted beam) and $\theta+180$\si{\degree} (reflected beam) resulting in an unshifted signal due to the reflection symmetry of the heading error. To demonstrate the difference in the signal between a double-pass and a split-beam, we assume a Bell-Bloom magnetometer\,\cite{bell1961optically}. We model the magnetic resonance using the Bloch equations
\begin{equation}\label{Eq: bloch}
    \dot{\vec{M}}=\gamma\vec{M}\times\vec{B}-\Gamma\vec{M}+R(t)\vec{M_0}\,,
\end{equation}
where $\vec{M}$ is the magnetization, and $\dot{\vec{M}}$ is its time derivative, $\gamma$ is the gyromagnetic ratio, $\vec{B}$ is the magnetic field, $\Gamma$ is the relaxation rate, $\vec{M_0}$ is the maximum polarization in the absence of relaxation and $R(t)$ is the time depended pumping rate. Assuming a pumping rate of $R(t)=\frac{R_o}{2}[1+\cos(\omega t)]$ along the $x$ axis and a magnetic field along the $z$ axis, we solve for the steady-state in the rotating frame under the rotating frame approximation and rotate the solution back to the lab frame to get
\begin{equation}\label{Eq: bloch Mx}
    M_x=\frac{1}{4}R_0M_0\frac{\Gamma\cos(\omega t)+(\omega-\omega_L)\sin(\omega t)}{(\omega-\omega_L)^2+\Gamma^2}\,,
\end{equation}
where $\omega_L=\gamma B_z$ is the Larmor frequency. The solution has a Lorentzian line shape component (the in-phase) and a dispersive line shape component (the quadrature). Typically, we extract the resonance by finding the zero-crossing of the quadrature as it has a larger response to a change in the magnetic field than the peak of the in-phase. In order to have a dispersive line shape in a split-beam configuration, one has to subtract the in-phase of each beam, while the double-pass effectively sums two quadratures. Summing two quadratures and subtracting two in-phase signals results in different line shapes for different heading error values. For example, if one beam has the same Larmor frequency as the other beam (i.e., zero heading error), a subtraction of the signals results in a null signal. In contrast, a summation (as in double-pass) will have its maximum signal for such a case. See typical line shapes in Fig.\,\ref{fig: u and c shape}.
\par
\begin{figure}[ht]
\centering
\includegraphics[width=1\columnwidth]{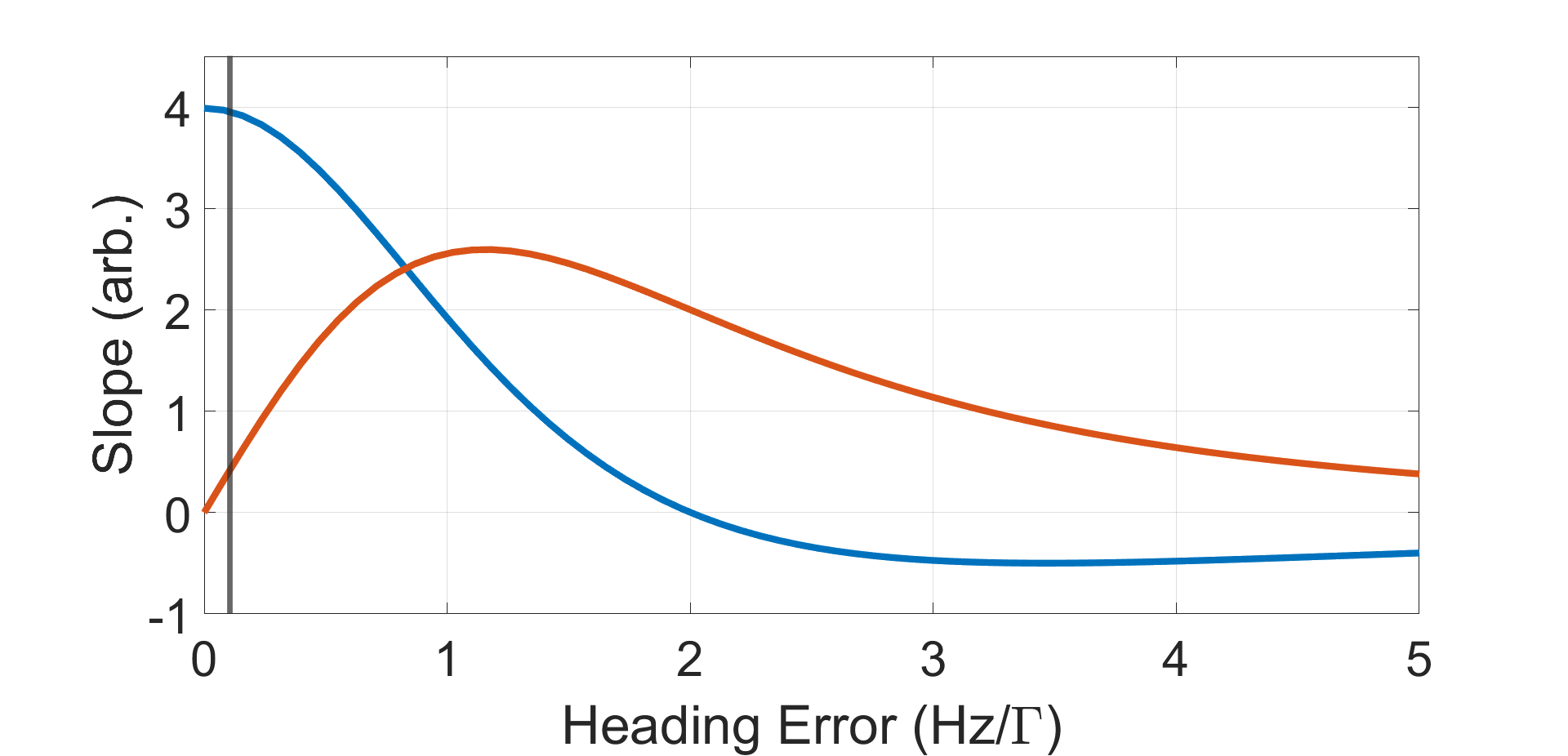}
\caption{The slope of the signal of a split-beam configuration (red) compared to the slope of the double-pass configuration (blue) as a function of heading error in units of relaxation rate (i.e., the width of the magnetic resonance line shape). The negative values in the blue line, which might seem unnatural given that the signal is a summation of two line shapes with a positive slope (see Fig.\,\ref{fig: u and c shape}), are due to a deformation near the center when the two quadrature signals are separated more than two $\Gamma$. The black line near $x=0.1$ represents the typical $x$ value for the Earth magnetic field in a Cs magnetometer ($53\,$Hz heading error and $\Gamma=500\,$Hz). At this point, the double-pass configuration exhibits an order-of-magnitude improvement in the slope (amplitude of the magnetic resonance line shape signal vs. the modulation frequency as depicted in Fig.\,\ref{fig: u and c shape}), giving rise to an order-of-magnitude theoretical improved sensitivity, even without considering the increased interaction length of the double-pass, making the double-pass a preferable choice for Earth-field magnetometers.}
\label{fig: slope}
\end{figure}

In order to study the difference between the two methods, we calculate the slope at the center of the subtracted/summed signal, which is proportional to the sensitivity, as a function of the difference in the resonance frequency between the two signals (i.e., heading error). We can see in Fig.\,\ref{fig: slope} that for actual magnetometer values: $\Gamma=500\,$Hz and $50\,$nT heading error (i.e., $x=0.1$ in the figure), the signal that arises from the double-pass is much stronger than that of the split-beam.
\section{Experimental Set-Up and Results}\label{Sec: experimental}
\begin{figure}[t]
\centering
\includegraphics[width=\columnwidth]{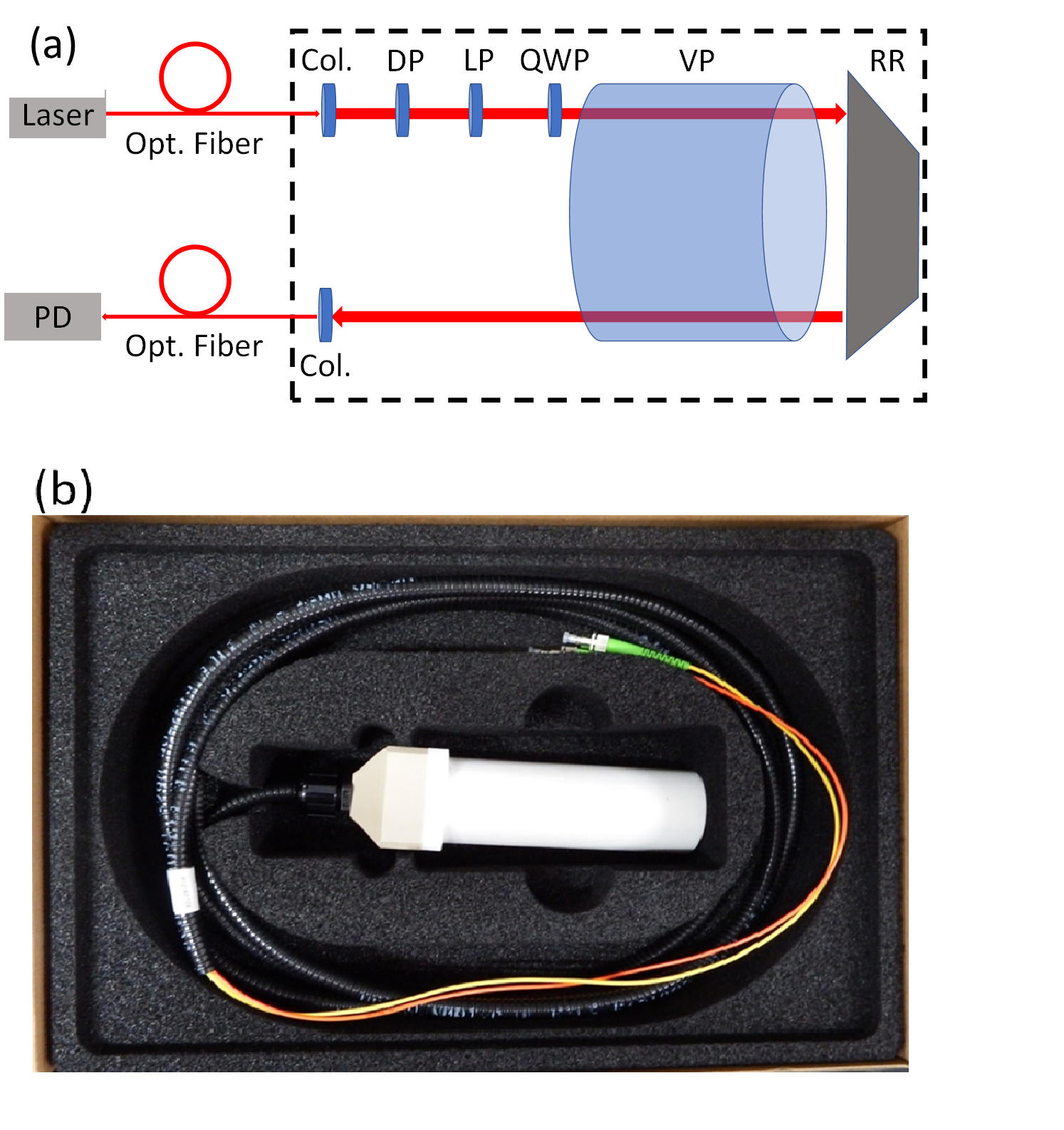}     
\caption{(a) Schematic diagram of the sensor head. A frequency-modulated laser (Laser) is coupled to a 10\,m optical fiber (Opt. Fiber) to separate the sensor head from the electronics. The end of the 10\,m fiber is coupled to a collimator (Col.) which expands the beam into free space. The free-space laser is circularly polarized using a depolarizer (DP), linear polarizer (LP), and a quarter wavelength plate (QWP). The DP is required because, due to the motion of the fiber in the portable platform, the polarization at the output of the fiber is fluctuating which induces excess noise. After the laser traverses the vapor cell (VP), it is retro-reflected (RR) back to the cell using a set of mirrors that keeps the helicity unchanged. The laser is then collimated back into a 10\,m fiber using a second collimator (Col.) and then directed into a photodiode (PD). (b) Picture of the sensor head, including the two 10\,m optical fibers. All of the sensor head parts are made from non-magnetic parts verified in our lab.}
\label{fig: actual}
\end{figure}
The double-pass sensor head is all-optical with no electronics inside, driven by a Vertical-Cavity Surface-Emitting Laser (VCSEL). We stabilize the VCSEL's wavelength using a current source and temperature controller. However, a drift in the environmental temperature of the laser can lead to a drift in the wavelength despite the stabilization. The drift in the temperate has a low-frequency component (mHz and below, while typical magnetic anomalies are 0.1\,Hz and above in a portable field platform\,\cite{wang2021frequency}), but once the wavelength is not at the optimal value the sensitivity in all frequencies is degraded. Therefore, to have a high-performance sensor in a portable platform, we add a 2\textsuperscript{nd} degree temperature controller that stabilizes the laser to an arbitrary environmental temperature. Another problem with transitioning the sensor head from the lab to a portable field platform is that the movement of the optical fiber connected to the sensor head induces changes in the laser's polarization at the optical fiber's output. Using a set of linear polarizers and a quarter-wavelength retarder to polarize the laser circularly will require frequent calibration, which is impossible in a field magnetometer. Alternatively, using electrically controlled polarizers or retarders will significantly raise the price of the magnetometer and, more importantly,  can add magnetic noise from its currents. Thus, we add a depolarizer after the optical fiber and before the linear polarizer as depicted in Fig.\,\ref{fig: actual}, eliminating the polarization fluctuation from the fiber movement (by depolarizing the light\,\cite{patentDP}) at the cost of laser power. We use a set of four mirrors to reflect the laser back to the cell with the same helicity. The reflection back to the cell can be done with two mirrors, oriented in 45\si{\degree}, but while the reflectance of typical dielectric mirrors is equal for both S and P polarization (e.g., Thorlabs BB mirrors), the phase shift is not equal for S and P polarization. The results of circularly polarized light reflected from two mirrors with different phase shifts for the different linear polarization is elliptical polarization which is unwanted. To balance the phase shift between S and P components of the light, we use a set of four of-the-shelf mirrors that compensate for the different phase shifts. Alternatively, one can use two specially designed mirrors. See the schematics of the double-pass sensor head and the actual portable sensor in Fig.\,\ref{fig: actual}.
\par
\begin{figure}[ht]
\centering
\includegraphics[width=0.45\textwidth]{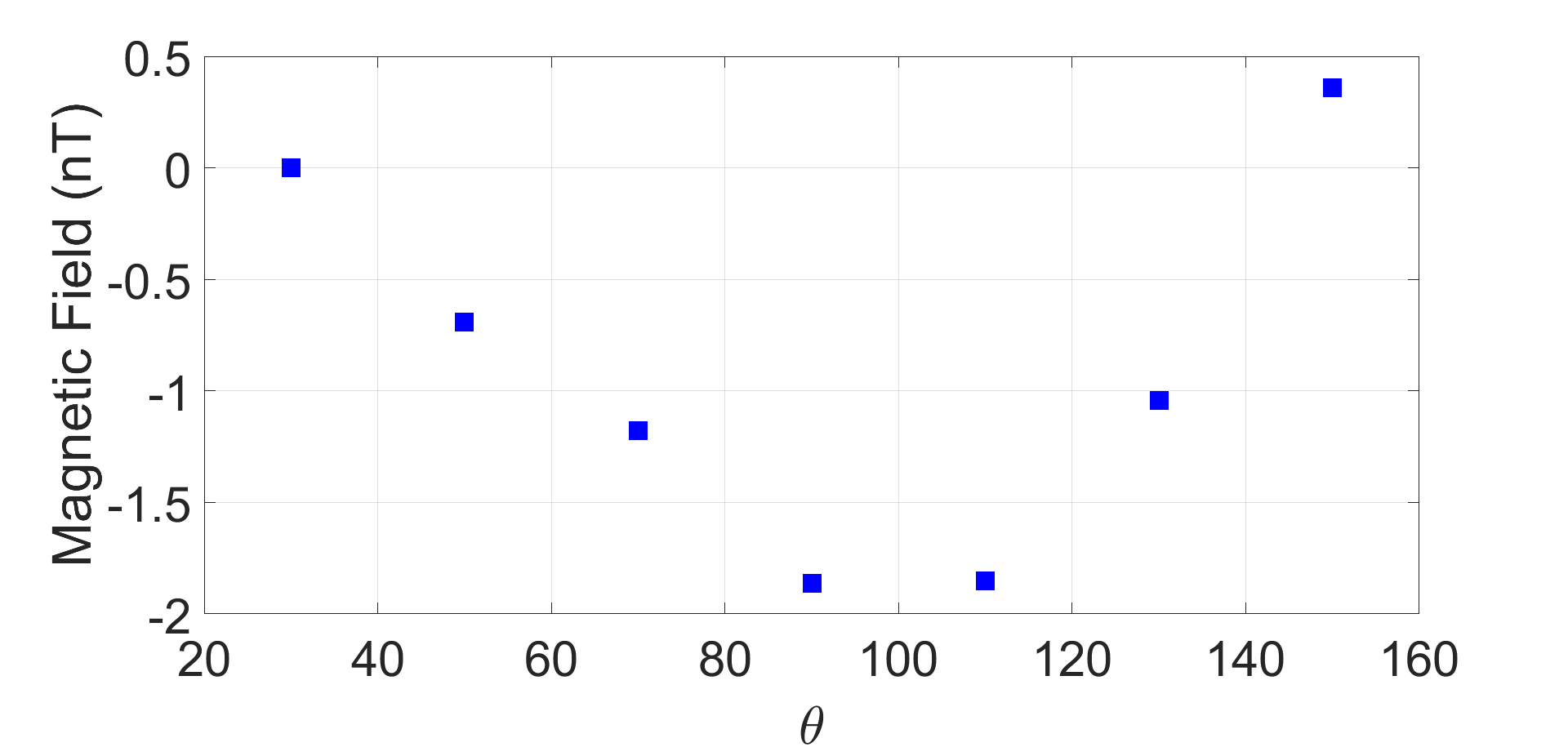}
\caption{Heading error experimental results. The maximal heading error is about 2\,nT, which is an order of magnitude improvement compared to the theoretical calculation presented above for Cs Earth-field magnetometers without heading error compensation and also performs much better than a commercial all-optical Cs magnetoemter\,\cite{MFAM} which shows a 10\,nT heading error in the best configuration. A 0.5\,nT noise contribution to the heading error presented above is attributed to the large curvature of the fiber at large angles, which was discovered only during the field experiment. Error bars are too small to be visible.}
\label{fig: he_results}
\end{figure}

 Heading error measurements were done by positioning two identical sensor heads (with a sensitivity of $\approx$4\,pT at 1\,Hz) 3\,m apart in a magnetically quiet area. One of the sensors serves as a reference sensor with $\theta=\ang{90}$, and the other measures the magnetic field at different $\theta$ angles. The measurement range is limited to $\ang{30}\leq \theta \geq \ang{150}$ due to the dead zone of the sensor as well as at larger/smaller angles, the fiber at the entrance to the sensor head has a significant curvature which induces noise larger than the heading error (in an actual operation this is not an issue as the sensor is mounted to the portable platform with minimal curvature). However, even within that range, we estimate the excess noise contribution to the heading error measurement due to the fiber curvature to be $\approx0.5$\,nT at large measurement angles. All the materials of the sensor head and the heading error setup were tested to be non-magnetic using a commercial sensor with $\approx$4\,pT at 1\,Hz sensitivity. The two sensors' magnetic field readouts are subtracted to eliminate errors related to Earth's magnetic field diurnal variations. The gradiometer reading is calculated at different angles and is shown in Fig\,\ref{fig: he_results}.

\begin{figure}[ht]
\centering
\includegraphics[width=0.4\textwidth]{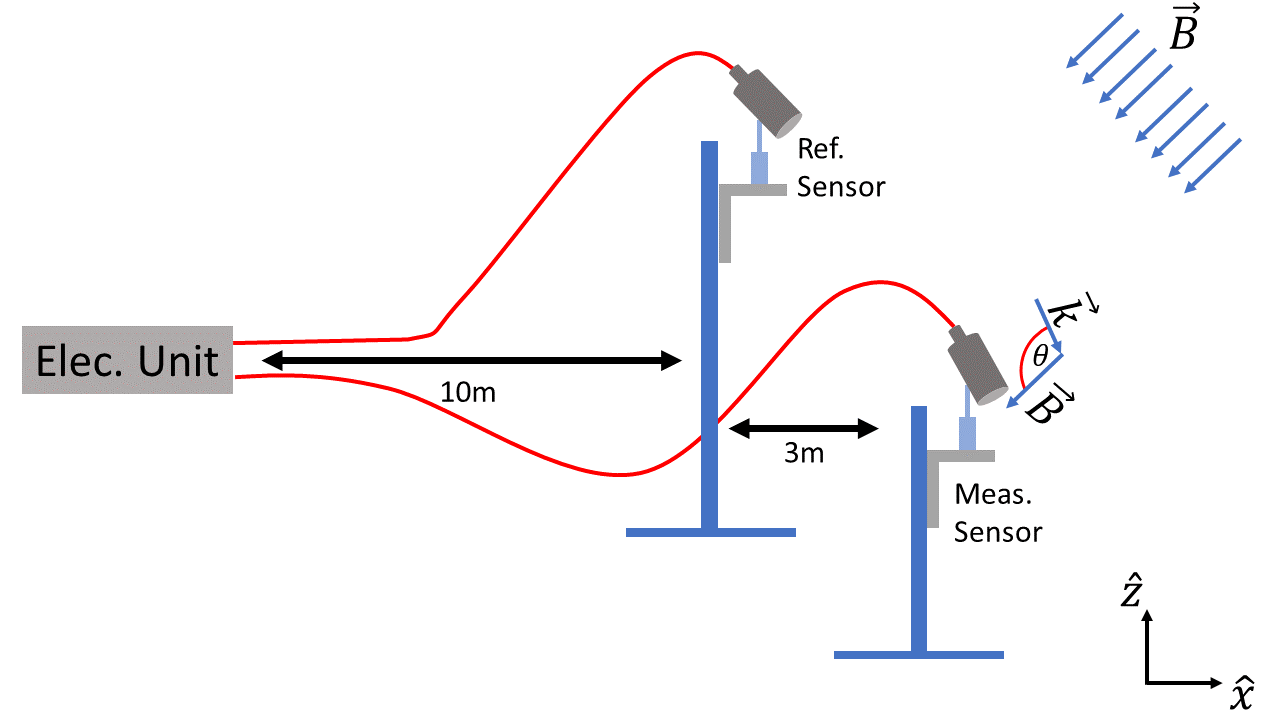}
\caption{Schematic diagram of the setup. A reference sensor (Ref. Sensor) and another sensor for the measurements (Meas. Sensor) are positioned 3\,m apart from each other and 10\,m away from the electronic unit (Elec. Unit). The angle $\theta$ between the measurement sensor optical axis ($\vec{k}$ arrow) and the magnetic field axis ($\vec{B}$ arrow) is changed in steps of $\ang{20}$ by rotating the sensor, while the reference sensor is kept at $\theta=\ang{90}$ throughout the experiment. The magnetic field readout of the two sensors is subtracted to reomve Earth's magnetic field diurnal variations and other distant magnetic targets. After each change of $\theta$, we measure the relative change in the gradiometer reading.}
\label{fig: setup}
\end{figure}
\section{Discussion and Conclusion}\label{Sec: conclusion}
The double-pass configuration heading error is estimated to be $\approx2$\,nT - an order of magnitude improvement compared to our theoretical estimation for a magnetometer without heading error compensation (and the estimation in \cite{hrvoic2005brief}) and better than in a modern commercial all-optical portable optically pumped magnetometer\,\cite{MFAM,comment}.
\par
The measurement unit is estimated to be responsible for a quarter of the 2\,nT heading error due to the fiber curvature at large angles and can be removed in an actual installation to a platform (e.g., by fixing the fiber pigtail when the sensor is attached to the mobile platform). In addition, the movement of the 10\,m fibers in a mobile platform introduces a noise, which was resolved by adding a depolarizer before the linear polarizer (see Fig.\,\ref{fig: setup}). However, introducing a depolarizer followed by a linear polarizer results in a significant power loss which can avoid by inserting the laser into the sensor head. 
\par
The level of the heading error compensation depends on balancing between the incoming and reflected beam in the double-pass configuration or between the two beams in the split-beam configuration. The split-beam configuration requires careful and frequent balancing and/or post-processing of the two signals\,\cite{oelsner2022integrated}. A double-pass configuration, which sums signals instead of subtracting them, is less sensitive to fluctuations in the magnetic resonance amplitudes, and these drawbacks do not appear. However, this advantage comes at the cost of an inherent laser power imbalance between the incoming and reflected beams. The difference in laser power translates into a difference in the magnetic resonance amplitude between the two beams and results in an uncompensated signal, which can explain the remaining heading error. Nevertheless, higher laser power, which can be achieved by placing the laser inside the sensor head, can help bypass this effect as the change in the magnetic resonance amplitude between the incoming and reflected beams becomes negligible at higher laser power. 
\par
Finally, the interaction length is increased due to the double-pass configuration and can enable the miniaturization of the sensor by reducing the vapor temperature and its associated spin-exchange relaxation rate. 
\par
This work opens the door for a simple and robust sub-pT portable sensor in Earth field\,\cite{patentHE}. 
\begin{acknowledgments}
We would like to thank Yonatan Japha and Tetyana Kuzmenko for useful discussions, and Elta's R\&D team, Ronen Wolf, Avi Elmalem, Gil Shalev, Eran Domb, Shahar Laykin and Ravid Avital for their support.
This work was funded in part by the Israeli Science Foundation Grants 1314/19, 3515/20 and by the Israeli Innovation Authority Grant No. 74482.
\end{acknowledgments}
\bibliographystyle{ieeetr}
\bibliography{sample.bib}

\begin{thebibliography}{10}

\bibitem{allred2002high}
J.~Allred, R.~Lyman, T.~Kornack, and M.~V. Romalis, ``High-sensitivity atomic
  magnetometer unaffected by spin-exchange relaxation,'' {\em Physical review
  letters}, vol.~89, no.~13, p.~130801, 2002.

\bibitem{dang2010ultrahigh}
H.~Dang, A.~C. Maloof, and M.~V. Romalis, ``Ultrahigh sensitivity magnetic
  field and magnetization measurements with an atomic magnetometer,'' {\em
  Applied Physics Letters}, vol.~97, no.~15, p.~151110, 2010.

\bibitem{sheng2013subfemtotesla}
D.~Sheng, S.~Li, N.~Dural, and M.~V. Romalis, ``Subfemtotesla scalar atomic
  magnetometry using multipass cells,'' {\em Physical review letters},
  vol.~110, no.~16, p.~160802, 2013.

\bibitem{pratt2021kernel}
E.~J. Pratt, M.~Ledbetter, R.~Jim{\'e}nez-Mart{\'\i}nez, B.~Shapiro, A.~Solon,
  G.~Z. Iwata, S.~Garber, J.~Gormley, D.~Decker, D.~Delgadillo, {\em et~al.},
  ``Kernel flux: a whole-head 432-magnetometer optically-pumped
  magnetoencephalography (op-meg) system for brain activity imaging during
  natural human experiences,'' in {\em Optical and Quantum Sensing and
  Precision Metrology}, vol.~11700, pp.~162--179, SPIE, 2021.

\bibitem{shah2013compact}
V.~K. Shah and R.~T. Wakai, ``A compact, high performance atomic magnetometer
  for biomedical applications,'' {\em Physics in Medicine \& Biology}, vol.~58,
  no.~22, p.~8153, 2013.

\bibitem{wang2018application}
T.~Wang, D.~F.~J. Kimball, A.~O. Sushkov, D.~Aybas, J.~W. Blanchard,
  G.~Centers, S.~R. O’Kelley, A.~Wickenbrock, J.~Fang, and D.~Budker,
  ``Application of spin-exchange relaxation-free magnetometry to the cosmic
  axion spin precession experiment,'' {\em Physics of the dark universe},
  vol.~19, pp.~27--35, 2018.

\bibitem{lucivero2014shot}
V.~G. Lucivero, P.~Anielski, W.~Gawlik, and M.~W. Mitchell,
  ``Shot-noise-limited magnetometer with sub-picotesla sensitivity at room
  temperature,'' {\em Review of Scientific Instruments}, vol.~85, no.~11,
  p.~113108, 2014.

\bibitem{acosta2006nonlinear}
V.~Acosta, M.~Ledbetter, S.~Rochester, D.~Budker, D.~J. Kimball, D.~Hovde,
  W.~Gawlik, S.~Pustelny, J.~Zachorowski, and V.~Yashchuk, ``Nonlinear
  magneto-optical rotation with frequency-modulated light in the geophysical
  field range,'' {\em Physical Review A}, vol.~73, no.~5, p.~053404, 2006.

\bibitem{oelsner2022integrated}
G.~Oelsner, R.~IJsselsteijn, T.~Scholtes, A.~Kr{\"u}ger, V.~Schultze,
  G.~Seyffert, G.~Werner, M.~J{\"a}ger, A.~Chwala, and R.~Stolz, ``Integrated
  optically pumped magnetometer for measurements within earth’s magnetic
  field,'' {\em Physical Review Applied}, vol.~17, no.~2, p.~024034, 2022.

\bibitem{limes2020portable}
M.~Limes, E.~Foley, T.~Kornack, S.~Caliga, S.~McBride, A.~Braun, W.~Lee,
  V.~Lucivero, and M.~Romalis, ``Portable magnetometry for detection of
  biomagnetism in ambient environments,'' {\em Physical Review Applied},
  vol.~14, no.~1, p.~011002, 2020.

\bibitem{li2016unshielded}
W.~Li, X.~Peng, S.~Li, C.~Liu, H.~Guo, P.~Lin, and W.~Zhang, ``Unshielded
  scalar magnetometer based on nonlinear magneto-optical rotation with
  amplitude modulated light,'' in {\em 2016 IEEE International Frequency
  Control Symposium (IFCS)}, pp.~1--4, IEEE, 2016.

\bibitem{zhang2023heading}
R.~Zhang, D.~Kanta, A.~Wickenbrock, H.~Guo, and D.~Budker, ``Heading-error-free
  optical atomic magnetometry in the earth-field range,'' {\em Physical Review
  Letters}, vol.~130, no.~15, p.~153601, 2023.

\bibitem{schultze2017optically}
V.~Schultze, B.~Schillig, R.~IJsselsteijn, T.~Scholtes, S.~Woetzel, and
  R.~Stolz, ``An optically pumped magnetometer working in the light-shift
  dispersed m z mode,'' {\em Sensors}, vol.~17, no.~3, p.~561, 2017.

\bibitem{bell1961optically}
W.~E. Bell and A.~L. Bloom, ``Optically driven spin precession,'' {\em Physical
  Review Letters}, vol.~6, no.~6, p.~280, 1961.

\bibitem{hrvoic2005brief}
I.~Hrvoic, G.~M. Hollyer, and P.~Eng, ``Brief review of quantum
  magnetometers,'' {\em GEM Systems Technical Papers}, 2005.

\bibitem{yabuzaki1974frequency}
T.~Yabuzaki and T.~Ogawa, ``Frequency shifts of self-oscillating magnetometer
  with cesium vapor,'' {\em Journal of Applied Physics}, vol.~45, no.~3,
  pp.~1342--1355, 1974.

\bibitem{seltzer2007synchronous}
S.~Seltzer, P.~Meares, and M.~Romalis, ``Synchronous optical pumping of quantum
  revival beats for atomic magnetometry,'' {\em Physical Review A}, vol.~75,
  no.~5, p.~051407, 2007.

\bibitem{bao2018suppression}
G.~Bao, A.~Wickenbrock, S.~Rochester, W.~Zhang, and D.~Budker, ``Suppression of
  the nonlinear zeeman effect and heading error in earth-field-range
  alkali-vapor magnetometers,'' {\em Physical Review Letters}, vol.~120, no.~3,
  p.~033202, 2018.

\bibitem{gem}
https://www.gemsys.ca/

\bibitem{scintrexltd}
https://scintrexltd.com/applications/magnetics//

\bibitem{hovde2013commercial}
D.~Hovde, M.~Prouty, I.~Hrvoic, and R.~Slocum, ``Commercial magnetometers and
  their application,'' {\em Optical Magnetometry}, vol.~2013, pp.~387--405,
  2013.

\bibitem{oelsner2019sources}
G.~Oelsner, V.~Schultze, R.~IJsselsteijn, F.~Wittk{\"a}mper, and R.~Stolz,
  ``Sources of heading errors in optically pumped magnetometers operated in the
  earth's magnetic field,'' {\em Physical Review A}, vol.~99, no.~1, p.~013420,
  2019.

\bibitem{steck2007quantum}
D.~A. Steck, ``Quantum and atom optics,'' 2007.

\bibitem{kimball_2013}
D.~Budker and D.~Kimball, {\em Optical Magnetometry}.
\newblock Cambridge University Press, 2013.

\bibitem{wang2021frequency}
J.~Wang, Z.~Jiang, J.~Gao, S.~Zhao, W.~Zhai, and Y.~Shen, ``Frequency
  characteristics analysis for magnetic anomaly detection,'' {\em IEEE
  Geoscience and Remote Sensing Letters}, vol.~19, pp.~1--5, 2021.

\bibitem{patentDP}
I.~Shcherback {\em et~al.}, ``A magnetometer and methods,'' Feb.~2 2022.
\newblock IL Patent 290465.

\bibitem{MFAM}
https://www.geometrics.com/product/mfam-developer kit/

\bibitem{comment}
 It is a tricky task to compare magnetometers as there is no such thing as the
  best magnetometer - there is a magnetometer that is best suited for your
  needs. Having said that, we would like to add some more insights to our
  comparison. Our statement at the end of the first paragraph of Sec. IV, is
  based on a comparison of our all-optical Cs-based magnetometer to a similar
  cutting-edge commercial product (MFAM, Ref. 26) which has 10\,nT heading
  error while we measure 2\,nT. We also measure 4\,pT/$\sqrt{\mathrm{Hz}}$ at
  1\,Hz which is roughly the same as in the case of the MFAM. Another advantage
  of our system is that while the MFAM uses two separate lasers for pumping and
  probing, we use a single laser which is beneficial in terms of cost-of-goods,
  power budget, complexity, etc. Other top-level commercial magnetometers, such
  as GSMP-35u by GEM systems or QTFM by Quspin are not all optical. The GSMP
  has an outstanding sensitivity of 0.2\,pT at 1Hz, and better heading error
  due to the narrow spectral line of the K vapor. But it comes at the cost of a
  very low band-width (maximal 20\,Hz sample rate) and the lack of laser
  availability at the K frequency. Moreover, The GSMP also has polar and
  equatorial dead zones compared to the single equatorial dead zones in our
  Bell-bloom magnetometer. The QTFM of Quspin has similar performances to our
  magnetometer with 3\,pT sensitivity and 3\,nT heading error but requires a
  strong holding field during the pumping time, which adds complication to the
  setup compared to an all-optical Bell-Bloom magnetometer.

\bibitem{patentHE}
Y.~Rosenzweig {\em et~al.}, ``Optically pumped magnetometer,'' June~29 2016.
\newblock IL Patent 304165.

\end{thebibliography}
\end{document}